\documentclass[12pt]{article}
\usepackage{amssymb,latexsym,epsfig,cite}

\newcommand{\be}{\begin{equation}}
\newcommand{\ee}{\end{equation}}
\def\bea{\begin{eqnarray}}
\def\eea{\end{eqnarray}}
\begin{document}

\thispagestyle{empty}

\begin{center}
{\Large Optical potentials using resonance states in\\[1.5ex] 
Supersymmetric Quantum Mechanics}
\end{center}

\vskip0.1mm

\begin{center}
Nicol\'as Fern\'andez-Garc\'{\i}a and Oscar Rosas-Ortiz\\[2ex]
{\footnotesize Departamento de F\'{\i}sica, Cinvestav, AP 14-740, 07000
M\'exico DF, Mexico}
\end{center}

\begin{abstract}
Complex potentials are constructed as Darboux-deformations of short range,
radial nonsingular potentials. They behave as optical devices which both
refracts and absorbs light waves. The deformation preserves the initial
spectrum of energies and it is implemented by means of a Gamow-Siegert
function (resonance state). As straightforward example, the method is
applied to the radial square well. Analytical derivations of the involved
resonances show that they are `quantized' while the corresponding
wave-functions are shown to behave as bounded states under the broken of
parity symmetry of the related one-dimensional problem. 
\end{abstract}

\section{Introduction}

Solutions of the Schr\"odinger equation associated to complex eigenvalues
$\epsilon = E_{\alpha}$ and satisfying purely outgoing conditions are
known as Gamow-Siegert functions \cite{Gam28,Sie39}. These solutions
represent a special case of scattering states for which the `capture' of
the incident wave produces delays in the scattered wave. The `time of
capture' can be connected with the lifetime of a decaying system
(resonance state) which is composed by the scatterer and the incident
wave. Then, it is usual to take ${\rm Re}(\epsilon)$ as the binding energy
of the composite while ${\rm Im}(\epsilon)$ corresponds to the inverse of
its lifetime. The Gamow-Siegert functions are not admissible as physical
solutions into the mathematical structure of Quantum Mechanics since, in
contrast with conventional scattering wave-functions, they are not finite
at $r\rightarrow \infty$. Thus, such a kind of functions is acceptable in
Quantum Mechanics only as a convenient model to solve scattering
equations. However, because of the resonance states relevance, some
approaches extend the formalism of quantum theory so that they can be
defined in a precise form
\cite{Boh89,Boh89_2,del02,Civ04,Agu71,Bal71,Sim72}.

In this paper the Gamow-Siegert functions are analyzed not to precisely
represent decaying systems but to be used as the cornerstone of complex
Darboux-transformations in Supersymmetric Quantum Mechanics (Susy-QM). The
`unphysical' behaviour of these solutions plays a relevant role in the
transformation: While the Gamow-Siegert function $u$ diverges at $r
\rightarrow \infty$, the limit of its logarithmic derivative $\beta =
-u'/u$ at $r \rightarrow \infty$ is a complex constant. This last function
is used to deform the initial potential into a complex function. It is
worth noticing that complex eigenvalues $\epsilon$ have been used as
factorization constants in Susy-QM (see for instance
\cite{Can98,And99,Bag01,Fer03,Ros03,Sam06a,Sam06b,Sam06c,Cab07} and the
discussion on `atypical models' in \cite{Mie04}). However, as far as we
know, until the recent results reported in \cite{Ros07,Fer07a,Fer07b} the
connection between Susy-QM and resonance states has been missing. Our
interest in the present work is two-fold. First, we want to show that
appropriate approximations lead to analytical expressions for the real and
imaginary parts of the resonance energies in the case of radial short
range potentials. Second, we want to enlarge the number of applications of
the Darboux transformations in constructing new exactly solvable models in
Quantum Mechanics.

In Section~2 the main aspects of solving the Schr\"odinger equation for
radial, nonsingular short range potentials are reviewed. The relevance of
the scattering amplitude $S(k)$ in the analysis of both the bounded and
scattering wave-functions is clearly stated. In Section~3 the analytical
properties of $S(k)$ as a function of the complex kinetic parameter $k$
are studied. It is shown that poles of the scattering amplitude which live
in the lower $k$--plane are connected with the Gamow-Siegert functions
while the poles on the positive imaginary axis lead to bounded physical
energies. The general aspects of the complex Darboux-transformations are
included in Section~4. All the previous developments are then applied to
the square radial well and new analytical expressions for the involved
resonance energies are reported in Section~5. We conclude the paper with
some concluding remarks in Section~6.

\section{Short range radial potentials revisited}
\subsection{General considerations}

Let us consider the Hamiltonian of one particle in the external,
spherically symmetric field $U(\vec r) = U(r)$. The Schr\"odinger
equation reduces to the eigenvalue problem: 
\be 
H \psi (\vec r) \equiv \left(-\frac{\hbar^2}{2m} \Delta + U(r) \right)
\psi (\vec r) = {\cal E} \psi (\vec r). 
\label{eigen1}
\ee 
We are interested in potentials which decrease more rapidly than $1/r$ at
large $r$. Indeed, $U(r)$ is a short-range potential (i.e., there exists
$a \in {\rm Dom} (U):= D_U$ such that $U(r) =0$ at $r>a$) which is
nonsingular (i.e., $U(r)$ satisfies $r^2 U(r) \rightarrow 0$ at $r
\rightarrow 0$). The wave function $\psi(\vec r)$ must be single-valued
and continuous everywhere in $D_U$. Since the operators $H$, $L^2$, $L_3$
and $P$ commute with each other, in spherical coordinates the
Schr\"odinger equation (\ref{eigen1}) has solutions of the form
\be 
\psi_{\ell m} (\vec r) = R_{\ell}(r) Y_{\ell m} (\theta, \varphi)
\label{sol1}
\ee 
where $\theta$ and $\varphi$ are respectively the polar and the azimuthal
angles, $R_{\ell}(r)$ is a function depending on $r$ and $Y_{\ell m}
(\theta, \varphi)$ stands for the spherical harmonics. For simplicity in
calculations we shall use the function $u_{\ell} (r) = r R_{\ell} (r)$.
The introduction of (\ref{sol1}) into equation (\ref{eigen1}) yields:
\be 
\left[ -\frac{d^2}{dr^2} + \frac{\ell(\ell+1)}{r^2} + v(r) \right]
u_{\ell}(r):= \left[ -\frac{d^2}{dr^2} + V_{\ell}(r) \right] u_{\ell}(r) =
k^2 u_{\ell}(r)
\label{schro1} 
\ee 
with $v = 2mU/\hbar^2$ and $k^2 =E \equiv 2m {\cal E}/\hbar^2$ the
dimensionless expressions for the potential and the kinetic parameter
respectively. As usual, the effective potential $V_{\ell}$ is defined as
the potential $v$ plus the centrifugal barrier $\ell (\ell +1)/r^2$.
Physical solutions $\psi (\vec r)$ of (\ref{eigen1}) should be constructed
with functions $u_\ell$ fulfilling the following boundary conditions:
\be
\left\{
\begin{array}{rl}
u_{\ell}(r) = 0 & r \rightarrow 0\\
\\
\frac{u_{\ell}(r)}{r} < \infty & r \rightarrow \infty
\end{array}
\right. 
\label{bound} 
\ee 
Furthermore, the function $u_{\ell}$ and its derivative $u_{\ell}'$ have
to be continuous for $r>0$ since (\ref{schro1}) includes second order
derivatives. As we shall see, the study of the involved matching
conditions is simplified by introducing a function $\beta_{\ell}$ as
follows
\be
\beta_{\ell}(r):= -\frac{u'_{\ell}(r)}{u_{\ell}(r)} = -\frac{d}{dr} \ln
u_{\ell}(r).
\label{beta} 
\ee 
In general we will assume that the energy spectrum is composed by negative
(bounded) and positive (scattering) energies. For the sake of notation we
shall use the symbol $\phi_{\ell}(r)$ for the solutions of (\ref{schro1})
which satisfy the boundary conditions (\ref{bound}); the symbol
$u_{\ell}(r)$ will stand for general mathematical solutions of equation
(\ref{schro1}).

\subsection{Bases of solutions}

For arbitrary $\ell$ and $r>a$, the appropriate basis of solutions
can be written in the form 
\be 
u_{\ell}^{(+)}(r) = ikr \,
h^{(1)}_{\ell}(kr), \qquad u_{\ell}^{(-)}(r) = -ikr \,
h^{(2)}_{\ell}(kr) 
\label{bessel2} 
\ee 
where $h^{(1)}_{\nu}$ and $h^{(2)}_{\nu}$ are the spherical H\"ankel
functions of order $\nu = \ell$ (see e.g. \cite{Abr72}). For large values
of $kr$ in the region $r>a$, the functions (\ref{bessel2}) behave as
follows
\be
u_{\ell}^{(\pm)}(r) \approx \exp \left[ \pm i \left(kr - \frac{\ell
\pi}{2}\right) \right]. 
\label{jost1} 
\ee 
Thereby, $u_{\ell}^{(+)}$ represents a diverging wave and describes
particles moving with speed $\vartheta =k$ in all directions from the
origin. In turn, the function $u_{\ell}^{(-)}$ represents a converging
wave and describes particles moving with speed $\vartheta$ towards the
origin.

It is worth noticing that the Schr\"odinger equation (\ref{schro1}) is
invariant under the change $k \rightarrow -k$. Hence, if $u_{\ell}(r,k)$
is solution of (\ref{schro1}) for $E = k^2$, then $u_{\ell}(r,-k)$ is also
a solution for the same energy. However, these last functions must differ
at most in a constant factor because the solutions of (\ref{schro1}) have
to be single-valued. In contrast, up to a global phase, they interchange
their roles at $r \rightarrow \infty$, as indicated by the relationship
\be
u_{\ell}^{(\pm)}(r,k) = e^{i\pi \ell} \, u_{\ell}^{(\mp)}(r,-k),
\quad r\rightarrow \infty. 
\label{jost2} 
\ee 
On the other hand, the basis of solutions at $r<a$ may be written
\be 
u^{(1)}_{\ell}(r) =
\frac{2\ell +1}{\Lambda_{\ell}} \, kr \, j_{\ell}(kr), \qquad
u^{(2)}_{\ell}(r) = - \Lambda_{\ell} \, kr \, n_{\ell}(kr), \qquad
\Lambda_{\ell} := \frac{2^{-\ell}\, \sqrt{\pi}}{\Gamma(\ell +1/2)}
\label{internas} 
\ee 
where $j_{\ell}$ and $n_{\ell}$ are respectively the spherical Bessel and
Neumann functions of order $\ell$. For very small values of $r$ we have
\be
u^{(1)}_{\ell}(r\rightarrow 0) \approx (kr)^{\ell+1}, \qquad
u^{(2)}_{\ell}(r \rightarrow 0) \approx \frac{1}{(kr)^{\ell}}.
\label{internas2} 
\ee

\subsubsection{Bounded states.}

Imaginary values of the kinetic parameter $k = \pm i \sqrt{\vert E
\vert} \equiv \pm i \kappa$ correspond to negative values of the
energy $E = k^2 = -\kappa^2$. As usual, we shall consider $k$ to be
in the upper complex $k$--plane $I_+$. In this way
$u^{(-)}_{\ell}(r,\kappa)$ does not satisfy the boundary condition
at $r\rightarrow \infty$ and $u^{(2)}_{\ell}(r,\kappa)$ does not
fulfill the boundary condition at $r=0$. Hence, the physical
solutions should be constructed with $u^{(1)}_{\ell}(r,\kappa)$ and
$u^{(+)}_{\ell}(r,\kappa)$. We write: 
\be 
\phi_{\ell}(r,\kappa) = \left\{
\begin{array}{rl} \zeta_{\ell}(\kappa) \, u_{\ell}^{(1)}(r,\kappa) \,
g_{\ell}(r,\kappa)& r<R\\
\\
\xi_{\ell}(\kappa)\, u^{(+)}_{\ell}(r,\kappa) & r \geq R
\end{array}
\right. 
\label{ligado1} 
\ee 
where the intermediary function $g_{\ell}(r,\kappa)$ is equal to 1 at
$r=0$ and depends on the potential. The coefficients
$\zeta_{\ell}(\kappa)$ and $\xi_{\ell}(\kappa)$ are complex numbers while
the matching condition $\beta_{\ell}(a,\kappa) =\kappa$ corresponds to a
transcendental equation which is fulfilled for a set of discrete values of
the kinetic parameter $\kappa_n$, $n=0,1,2,\ldots$ Under these conditions
the functions $\phi_{\ell}(r,\kappa_n)$ are finite and the integral
\be 
\int_0^{+\infty} \vert \phi_{\ell}(r,\kappa_n)
\vert^2 dr 
\label{square1} 
\ee 
converges. Thereby, $\vert \phi_{\ell}(r,\kappa_n) \vert^2$ can be
identified as the involved Born's probability density.

\subsubsection{Scattering states.}

Positive energies $E=k^2$ correspond to real values of $k$ and both
solutions (\ref{jost1}) remain finite in $r \geq a$. As a consequence,
both of them are physically acceptable in this region. To analyze these
solutions first let us consider the case $\ell =0$ and, for the sake of
simplicity, let us drop this subindex from the functions $u_{\ell}(r)$.
The general solution in the free of interaction zone ($r >a$) may be
written as
\be 
u_{\rm out} (r) =
\gamma(k)\,  \left[u^{(-)}(r) -S(k) u^{(+)}(r) \right].
\label{scat1} 
\ee 
The constant $\gamma(k)$ is the amplitude of the converging wave and does
not depend on the potential $v(r)$. The scattering amplitude $S(k)$, in
contrast, strongly depends on the potential and encodes the information of
the scattering phenomenon. The matching condition between $u_{\rm out}
(r)$ and the solution $u_{\rm in}(r)$ in the interaction zone reads
\be 
\beta_{\rm out}(a)
= \beta_{\rm in}(a) 
\label{out2} 
\ee 
and can be satisfied by appropriate values of $\gamma$ and $S$. Indeed,
equation (\ref{out2}) is equivalent to the following matrix array
\be 
\left.
\left(
\begin{array}{cc}
u_{\rm in} & u^{(+)}\\[1ex]
u'_{\rm in} & {u'}^{(+)}
\end{array}
\right) \right \vert_{r=a} \, \left(
\begin{array}{c}
\gamma^{-1}\\[1ex]
S
\end{array}
\right) =  \left. \left(
\begin{array}{c}
u^{(-)}\\[1ex]
{u'}^{(-)}
\end{array}
\right) \right\vert_{r=a} 
\label{matrix1} 
\ee 
the solution of which is given by the system of equations
\be 
\gamma= \left. \frac{W
\left( u_{\rm in}, u^{(+)} \right)}{2ik} \, \right\vert_{r=a},
\qquad S = \left. \frac{W \left( u_{\rm in}, u^{(-)}\right)}{W
\left( u_{\rm in}, u^{(+)} \right)} \, \right\vert_{r=a},
\label{sgamma} 
\ee 
where $W(\cdot,\cdot)$ stands for the Wronskian of the involved functions.

To get a better idea of the roles played by each one of the terms in
equation (\ref{scat1}) let us consider the free motion ($v(r) =0$) in the
whole of $D_v$. From equations (\ref{bessel2}) and (\ref{internas}) we
realize that a regular solution at the origin may be written as follows
\be 
u_{\rm free}(r) \equiv u^{(-)}(r) -
u^{(+)}(r) = -2ikr \, j_0(kr) = -2i \sin (kr). 
\label{close} 
\ee
This term can be introduced into equation (\ref{scat1}) to get 
\be
u_{\rm out} (r) = \gamma(k) \left[u_{\rm free}(r)-(S(k) -1)
u^{(+)}(r) \right]. 
\label{scat2} 
\ee 
Thus, the total wave function after scattering $u_{\rm out}$ is given by
the wavepacket one would have if there were no scattering $\gamma u_{\rm
free}$ plus a scattered term $-\gamma (S-1) u^{(+)}$. This last example
shows the important role played by the scattering amplitude in the
analysis of positive energies.

On the other hand, the potential we are dealing with is neither a sink nor
a source of particles, then the condition $\vert S(k) \vert =1$ (elastic
scattering) is true and we have
\be S(k)=
e^{2i\delta(k)}. 
\label{fase} 
\ee 
The scattering phase $\delta (k)$ represents the difference in phase of
the outgoing parts. The situation is clear if one introduces (\ref{close})
and (\ref{fase}) into equation (\ref{scat2}) to get
\be 
u_{\rm out}(r \rightarrow
\infty) =  -2i \gamma(k) \, e^{i\delta(k)} \, \sin (kr + \delta).
\label{close2} 
\ee 
As we have indicated, the scattering states include $k \in \mathbb{R}$
since $E>0$. However, the function (\ref{scat1}) is also valid if $k$ is
one of the imaginary values of the kinetic parameter leading to the
bounded energies $E_n=-\kappa_n^2$. Indeed, this case provides further
information about the scattering amplitude. For $k_n=i\kappa_n$, the
function (\ref{scat1}) at $r \rightarrow \infty$ reads
\be 
u_{\rm out}(r\rightarrow \infty,\kappa_n) \sim -2i \gamma(i\kappa_n)
\left[ e^{\kappa_n r} -S(i\kappa_n) e^{-\kappa_n r} \right]. 
\label{polo0}
\ee 
The first term in square brackets never vanishes ($\kappa_n >0$), then the
coefficient $S$ must be singular at the complex point $k_n =i\kappa_n$. In
other words, the points $k_n$ are nothing but poles of the scattering
amplitude. Since these points are in the upper complex $k$--plane $I_+$,
we realize that $S(k)$ is a regular function in $I_+$ except at $k_n$,
$n=0,1,2,\ldots$ We shall use these results in the next sections.

Finally, similar expressions can be found for arbitrary azimuthal
quantum numbers $\ell$. A straightforward calculation gives 
\be
u_{\ell}(r) = \gamma_{\ell}(k) \left( u^{(-)}_{\ell}(r) -
S_{\ell}(k) u^{(+)}_{\ell}(r) \right), \qquad r>a 
\label{close3} 
\ee
the asymptotic behaviour of which reads 
\be 
u_{\ell}(r \rightarrow
\infty) \approx -2i\gamma_{\ell}(k) e^{i\delta_{\ell}(k)} \, \sin
\left(kr + \delta_{\ell}(k) - \frac{\ell \pi}{2} \right).
\label{close4} 
\ee 
The above discussed properties of the scattering amplitude $S(k)$ are
easily generalized to the case of arbitrary angular momentum
$S_{\ell}(k)$.

\section{Gamow-Siegert functions}

\subsection{Analytic continuation of the scattering amplitude
$S_{\ell}(k)$}

In the previous section we realized that $S_{\ell}$ is a regular function
in $I_+$ except at the points $k=i\kappa$ leading to bound states of the
energy. In order to get a well behaved scattering amplitude in the whole
of the complex $k$--plane, this function must be extended to be analytic
in $I_-$. With this aim, first let us construct an arbitrary linear
combination of the functions (\ref{jost1}), it reads
\be 
u_{\ell}(r) = \zeta_{\ell}(k)\,
u_{\ell}^{(-)}(r) - \xi_{\ell}(k) \,u_{\ell}^{(+)}(r) \equiv
\zeta_{\ell}(k)\left[ u_{\ell}^{(-)}(r) - S_{\ell}(k)
u_{\ell}^{(+)}(r) \right] . 
\label{lin1} 
\ee 
This function is regular at the origin if $u_{\ell}(r=0)=0$, then we
arrive at the following relationship
\be 
S_{\ell}(k) =
\frac{\xi_{\ell}(k)}{\zeta_{\ell}(k)} = \left.
\frac{u_{\ell}^{(-)}(r)}{u_{\ell}^{(+)}(r)} \right\vert_{r=0}.
\label{s1} 
\ee 
Now, after a change of sign in $k$, the coefficients of the single-valued
function (\ref{lin1}) should satisfy
\be 
-e^{i \pi \ell} \xi_{\ell}(-k) = \alpha \zeta_{\ell}(k), \qquad e^{i \pi
\ell} \zeta_{\ell}(-k) = - \alpha \xi_{\ell}(k) 
\label{pato} 
\ee
where $\alpha$ is a proportionality factor and we have used equation
(\ref{jost2}). In this way, the expression (\ref{s1}) leads to 
\be
S_{\ell}(k) \, S_{\ell}(-k) =1. 
\label{s2} 
\ee 
Let $k_i \in I_+$ be a zero of $S_{\ell}$. From equation (\ref{s2}) we
notice that the scattering amplitude is well behaved in $I_-$ except at
the point $-k_i$, for which $S_{\ell}(-k_i) \rightarrow \infty$. Thus,
$-k_i \in I_-$ is a pole of $S_{\ell}$.

\begin{figure}[ht]
\begin{center}
\includegraphics[height=.20\textheight]{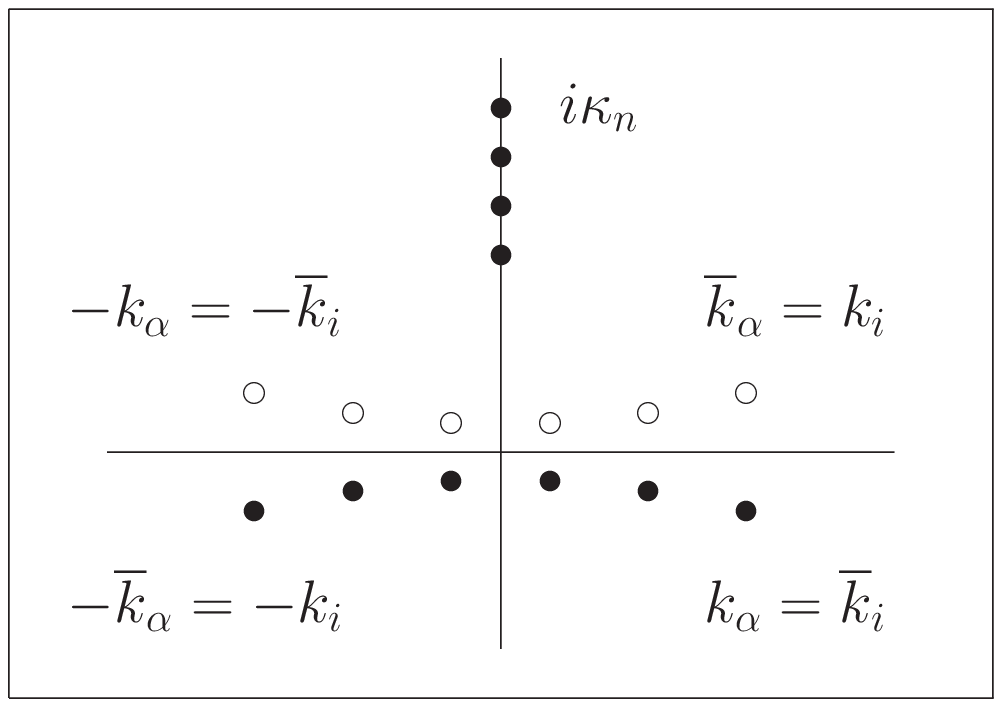}
\caption{\label{poles} \footnotesize Schematic representation of the poles
(disks) and the zeros (circles) of the scattering amplitude $S_{\ell}(k)$
in the complex $k$--plane. Bounded energies $E_n =-\kappa^2_n$ correspond
to the poles located on the positive imaginary axis.}
\end{center}
\end{figure}

On the other hand, if the kinetic parameter $k$ is real, then the complex
conjugate of any solution of equation (\ref{schro1}) is also a solution
for the same energy. However, the solution is unique so these last
functions differ at most in a global phase. Thereby, in a similar form as
in the previous case, we get
\be 
S_{\ell}(k) \, \overline S_{\ell}(k) =1
\label{s3} 
\ee 
where the bar stands for complex conjugation. This last equation is a
consequence of the elastic scattering we are dealing with since $\vert
S_{\ell}(k) \vert =1$. However, as $k$ is real, also equation (\ref{s3})
has to be extended to the whole of the complex $k$--plane. The natural
condition reads
\be 
S_{\ell}(k) \, \overline S_{\ell}(\overline k)
=1. 
\label{s4} 
\ee 
It is clear that $\overline S_{\ell}(\overline k_i)$ diverges because $k_i
\in I_+$ is a zero of $S_{\ell}$. In other words, $\overline k_i \in I_-$
is a pole of $S_{\ell}$ since $\vert S_{\ell} (\overline k_i) \vert
\rightarrow \infty$. Moreover, as $-k_i$ is a pole of $S_{\ell}$, from
equation (\ref{s4}) we also notice that $\overline S_{\ell}(-\overline
k_i)=0$. Thus, $-\overline k_i$ is another zero of the scattering
amplitude. In summary, if $k_{\alpha}$ ($k_i$) is a pole (zero) of
$S_{\ell}$, then $-\overline k_{\alpha}$ ($-\overline k_i$) is another
pole (zero) while $\overline k_{\alpha}$ and $-k_{\alpha}$ ($\overline
k_i$ and $-k_i$) are zeros (poles) of the scattering amplitude. In this
way $S_{\ell}(k)$ is a meromorphic function of the complex kinetic
parameter $k$, with poles restricted to the positive imaginary axis (bound
states) and to the lower half-plane $I_-$ (see Figure~\ref{poles}).

\subsection{Resonance states}

Let us express one of the poles of the scattering amplitude as $k_{\alpha}
= \alpha_1 - i \alpha_2$, with $\alpha_1$ an arbitrary real number and
$\alpha_2>0$. The function (\ref{close3}), evaluated for $k_{\alpha}$ at
$r\rightarrow \infty$ reads
\be 
u_{\ell}(r \rightarrow \infty,k_{\alpha}) \approx
\gamma_{\ell}(k_{\alpha})
e^{-i(\alpha_1 r - \ell \pi/2)} \left[ e^{-\alpha_2 r}
-S_{\ell}(k_{\alpha}) e^{\alpha_2 r} \right]. 
\label{gamo1} 
\ee
Observe that the first term in brackets vanishes for large values of
$-\alpha_2 r$. Hence, there remains only the scattering, purely outgoing
wave $e^{\alpha_2 r}$. As a consequence, the particle is maximally
scattered by the potential field $v(r)$ and the wave function is not
``well behaved'' since it does not fulfill the boundary condition at
$r\rightarrow \infty$. However, as we have seen, this last ``unphysical
behaviour'' is not only natural but necessary to study the elastic
scattering process of a particle by a short range, nonsingular radial
field interaction. Wave functions behaving like (\ref{gamo1}) are known as
Gamow-Siegert solutions of the Schr\"odinger equation (\ref{schro1}). We
write
\be
u^{(GS)}_{\ell}(r) = \left\{
\begin{array}{ll}
\theta_{\ell} \, g_{\ell}(r,k_{\alpha}) \,
u^{(1)}_{\ell}(k_{\alpha} r)& r<R\\
\\
-\gamma_{\ell}(k_{\alpha}) S_{\ell}(k_{\alpha})
u^{(+)}_{\ell}(r,k_{\alpha}) & R \leq r
\end{array}
\right. 
\label{gamosito} 
\ee 
where $\theta_{\ell}$ is a constant and the intermediary function
$g_{\ell}(r,k_{\alpha})$ is equal to 1 at $r=0$ and depends on the
potential. Besides the matching condition (\ref{sgamma}), the
Gamow-Siegert functions fulfill the purely outgoing boundary condition:
\be 
\lim\limits_{r \rightarrow+\infty}
\left\{ \beta^{(GS)}_{\ell}(r) + ik_{\alpha} \right\} =0.
\label{outgoing} 
\ee 
Our interest is now addressed to the Gamow-Siegert functions not precisely
as describing a decaying system but as appropriate mathematical tools to
study the construction of complex Darboux-deformed potentials.

\section{Complex-Darboux deformations of radial potentials}
\subsection{General considerations}

In order to throw further light on the function $\beta_{\ell}(r)$ we may
note that (\ref{beta}) transforms the Schr\"odinger equation
(\ref{schro1}) into a Riccati one
\be
-\beta_{\ell}'(r)+\beta^2_{\ell}(r)+\epsilon_{\ell} =V_{\ell}(r)
\label{riccati} 
\ee 
where the energy $k^2$ has been changed for the arbitrary number
$\epsilon_{\ell}$. Remark that (\ref{riccati}) is not invariant under a
change in the sign of the function beta:
\be
\beta_{\ell}'(r) + \beta^2_{\ell}(r) + \epsilon_{\ell}= V_{\ell}(r)
+ 2 \beta_{\ell}'(r). 
\label{nricatti} 
\ee 
These last equations define a Darboux transformation $\widetilde
V_{\ell}(r) \equiv \widetilde{V_{\ell}}(r, \epsilon_{\ell})$ of the
initial potential $V_{\ell}(r)$. If $\epsilon_{\ell}$ is a nontrivial
complex number, then the Darboux-deformation is necessarily a complex
function
\be
\widetilde V_{\ell}(r) = V_{\ell}(r) +2 \beta_{\ell}'(r) \equiv
V_{\ell}(r) - 2\frac{d^2}{dr^2} \ln \varphi_{\epsilon}(r)
\label{Vtilde} 
\ee 
where the {\rm transformation function} $u_{\ell}(r,\epsilon) \equiv
\varphi_{\epsilon}(r)$ is the general solution of (\ref{schro1}) for the
complex eigenvalue $\epsilon_{\ell}$. The solutions $y_{\ell}(r)=y_{\ell}
(r,\epsilon_{\ell},{\cal E})$ of the non-Hermitian Schr\"odinger equation
\be 
-y_{\ell}''(r)+\widetilde V_{\ell}(r) \, y_{\ell}(r)
={\cal E} y_{\ell}(r) 
\label{Schrod2} 
\ee 
are easily obtained 
\be
y_{\ell}(r) \propto \frac{{\rm W}(\varphi_{\epsilon}(r),
u_{\ell}(r))}{\varphi_{\epsilon}(r)}, 
\label{1ssol} 
\ee 
where $u_{\ell}(r)$ is eigensolution of equation (\ref{schro1}) with
eigenvalue ${\cal E}$.

\subsection{New exactly solvable optical potentials}

Let the transformation function $\varphi_{\epsilon}(r)$ be a Gamow-Siegert
function (\ref{gamosito}). The {\em superpotential} $\beta^{({\rm
GS})}_{\ell}(r,k_{\alpha})$ and the new potential $\widetilde V_{\ell}(r)$
behave at the edges of $D_v$ as follows
\be
\beta^{({\rm GS})}_{\ell}(r,k_{\alpha})= \left\{
\begin{array}{ll}
-\frac{\ell +1}{r} & r \rightarrow 0 \\[1ex]
-ik_{\alpha} & r \rightarrow +\infty
\end{array}
\right. \qquad \Rightarrow \qquad \widetilde V_{\ell}(r)= \left\{
\begin{array}{ll}
V_{\ell +1}(r) & r \rightarrow 0 \\[1ex]
V_{\ell}(r) & r \rightarrow +\infty
\end{array}
\right. 
\label{optical} 
\ee 
It is remarkable that even for an initial potential with $\ell =0$ the
corresponding Darboux-deformation contains a nontrivial centrifugal term
in the interaction zone $r<a$. Notice also that the new potential is
mainly a real function at the edges of $D_v$. Thus, the function ${\rm Im}
(\widetilde V_{\ell})$ is zero at the origin and vanishes at $r\rightarrow
+\infty$. In general ${\rm Im} (\widetilde V_{\ell})$ oscillates along the
whole of $(0,+\infty)$ according with the level of excitation of the
Gamow-Siegert solution $u_{\ell}^{({\rm GS})}(r,k_{\alpha})$: The higher
the level of excitation the greater the number of oscillations. Such a
behaviour implies a series of maxima and minima in ${\rm Im} (\widetilde
V_{\ell})$ which can be analyzed in terms of the optical bench
\cite{Mie04}.

As regards the physical solutions of the new potential it is immediate to
verify that, in all cases (bounded and scattering states of the initial
potential), the corresponding transformations fulfill the boundary
condition at $r=0$. The analysis of the behaviour at $r \rightarrow
\infty$ is as follows.

\subsubsection{Scattering states.}

Let $u_{\ell}(r)$ be a scattering state of the initial potential in $r>a$.
At $r\rightarrow +\infty$, the corresponding Darboux-Deformation
(\ref{1ssol}) behaves as follows
\be
\lim\limits_{r\rightarrow +\infty} y_{\ell}(r) = -i\gamma_{\ell}(k)
\left[ (k_{\alpha}+k) u^{(-)}(r) -(k_{\alpha}-k) S_{\ell}(k)
u^{(+)}(r) \right]_{r\rightarrow +\infty} 
\label{scater} 
\ee 
where we have used equation (\ref{close3}). As we can see, the
Darboux-deformations of $u_{\ell}$ behave as scattering states at $r>a$.

If the scattering state $u_{\ell}(r)$ is a Gamow-Siegert function and
$k\neq k_{\alpha}, \overline k_{\alpha}, -k_{\alpha}, - \overline
k_{\alpha}$, the Darboux-deformation preserves the purely outgoing
condition (\ref{outgoing}). Special cases are:
\begin{enumerate}
\item
$k=k_{\alpha} \Rightarrow$ the Wronskian in (\ref{1ssol}) is zero
and there is no transformation.

\item
$k=\overline k_{\alpha} \Rightarrow S_{\ell}(\overline
k_{\alpha})=0$, then $y_{\ell}(r)$ is an exponentially decreasing
function.

\item
$k= -k_{\alpha}  \Rightarrow S_{\ell}(-k_{\alpha})=0$ and the
coefficient of $u^{(-)}(r)$ vanishes, then $y_{\ell}(r)=0$.

\item
$k= -\overline k_{\alpha} \Rightarrow S_{\ell}(-\overline
k_{\alpha})\rightarrow \infty$, then $y_{\ell}(r)$ fulfills the
purely outgoing condition (\ref{outgoing}).

\end{enumerate}

\subsubsection{Bounded states.}

If $I_+ \ni k=i\kappa = i\sqrt{\vert E \vert}$ the function $u_{\ell} (r,
\sqrt{\vert E \vert})$ corresponds to one of the bounded states
(\ref{ligado1}). Thereby, from equation (\ref{1ssol}) we get
\be 
\lim\limits_{r\rightarrow +\infty} y_{\ell}(r,\sqrt{\vert
E \vert}) \approx (ik_{\alpha} - \sqrt{\vert E \vert})\,
\xi_{\ell}(\sqrt{\vert E \vert})\, e^{-i\frac{\ell
\pi}{2}}\,e^{-\sqrt{\vert E \vert}\,r}. 
\label{fisico1} 
\ee 
Thus, the Darboux-deformations of bounded wave-functions vanishes at
$r\rightarrow \infty$. However, although this new functions are
square-integrable, they do not form an orthogonal set in the Hilbert space
spanned by the initial square-integrable wave-functions \cite{Ros03} (see
also \cite{Ram03} and the `puzzles' with self orthogonal states
\cite{Sok06}).

In the next section we are going to solve the Schr\"odinger equation for
the radial square well. We shall get the solutions corresponding to
complex energy eigenvalues and purely outgoing boundary conditions. Thus,
we shall construct the involved Gamow-Siegert functions.

\section{Gamow-Siegert states of the radial square well}

Let us consider the potential 
\be 
v(r) = \left\{
\begin{array}{ll}
-v_0 & r<a\\[2ex]
0 & a \leq r
\end{array}
\right. 
\label{square} 
\ee 
with $a$ and $v_0$ positive real numbers. The regular solution of the
Schr\"odinger equation (\ref{schro1}) for this potential may be written
\be 
u_{\ell}(r) = \left\{
\begin{array}{cl}
\theta qr \, j_{\ell}(qr) & r<a\\[2ex]
-i\gamma_{\ell}(k) \, kr \left[h^{(2)}_{\ell}(kr) +S_{\ell}(k)
h^{(1)}_{\ell}(kr) \right] & a \leq r
\end{array}
\right. 
\label{gam0} 
\ee 
where the {\em interaction parameter} $q$ is defined by $q^2=v_0 +k^2$.
For simplicity, we shall analyze the $s$-wave solutions ($\ell =0$) for
which the matching conditions (\ref{sgamma}) lead to
\be 
S(q)=-\left[\frac{ik \sin (qa) + q \cos
(qa)}{ik \sin (qa) - q \cos (qa)}\right] e^{-2ika}, \qquad \gamma(k)
= -\frac{\theta e^{ika}}{2ik} [ik \sin (qa) - q \cos (qa)].
\label{match1} 
\ee 
Let us take $\theta=2ik$. The scattering amplitude $S(k)$ has zeros and
poles which respectively correspond to the roots of the transcendental
equations
\be 
\frac{iq}{k} = \tan
(qa), \qquad -\frac{iq}{k}= \tan (qa). 
\label{zeros} 
\ee 
Indeed, equations (\ref{zeros}) correspond to the even and odd
quantization conditions for negative energies if $k=i\kappa \in I_+$, just
as it has been discussed in the previous sections. However, we are looking
for points $k_{\alpha}$ in the complex $k$--plane such that ${\rm
Re}(k_{\alpha}) \neq 0$ and ${\rm Im}(k_{\alpha}) \neq 0$. Thereby, if
$k_{\alpha}$ is a pole of $S$, the converging wave in (\ref{gam0})
vanishes. Then the Gamow-Siegert function reads
\be
u^{({\rm GS})}_{\ell=0}(r) =\left\{
\begin{array}{cl}
2ik_{\alpha} \sin [q(k_{\alpha}) r] & r<a\\[2ex]
2ik_{\alpha} \sin [q(k_{\alpha}) a] \, e^{ik_{\alpha}(r-a)}& a \leq
r
\end{array}
\right. 
\label{gam1} 
\ee 
In order to solve the second equation in (\ref{zeros}) let us rewrite the
scattering amplitude as
\be 
S(k) =
- \left[ \frac{e^{2iqa}(k+q) +(q-k)}{e^{2iqa}(k-q) -(q+k)}\right]
e^{-2ika}. 
\label{match2} 
\ee 
In this way, the poles of $S$ are also the solutions of the transcendental
equation
\be 
e^{2iqa} =
\frac{\vert k \vert^2 - \vert q \vert^2 + 2 {\rm
Re}(k\overline{q})}{\vert k - q \vert^2}. 
\label{match3} 
\ee 
If the interaction parameter $q$ is real, then $\sin(2qa)=0$ and we obtain
the following quantization rule for the energy
\be 
q_n = \frac{\pi
n}{2a} \quad \Rightarrow \quad E_n = \left( \frac{n\pi}{2a}
\right)^2 -v_0, \quad n=0,1,\ldots 
\label{quant1} 
\ee 
On the other hand, let $k$ and $q$ be the complex numbers $k = K_1 +i K_2$
and $q = Q_1 +iQ_2$. Then $q^2 = v_0 + \epsilon = (v_0 +\epsilon_1) +i
\epsilon_2$, with $\epsilon = k^2$. Hence we have
\be 
Q_1^2 -Q_2^2 =
v_0 + \epsilon_1, \qquad 2Q_1Q_2=\epsilon_2 
\label{epsilon} 
\ee 
and
\be 
K_1^2 -K_2^2 = \epsilon_1, \qquad 2K_1K_2=\epsilon_2.
\label{epsilona} 
\ee 

The imaginary part of equation (\ref{match3}) is $\sin(2Q_1 a)=0$ and we
obtain the following quantization rule
\be 
Q_{1,n} = \frac{\pi n}{2a}, \quad n=0,1,\ldots 
\label{quant2}
\ee 
Therefore, if $\vert Q_2 \vert <<1$, the first equation in (\ref{epsilon})
shows that equation (\ref{quant2}) leads to $\epsilon_1 \approx E_n$. The
second equation in (\ref{epsilon}), on the other hand, gives a value of
$\epsilon_2$ which is not necessarily equal to zero. In this way, we shall
study the poles of $S$ for which the complex points $\epsilon$ are such
that ${\rm Re}(\epsilon) \approx E_n$ and ${\rm Im}(\epsilon) \approx
-\frac{\Gamma}{2}$, with $0< \Gamma <<1$. The main problem is then to
approximate the adequate value of $\Gamma$ as connected with the
discreteness of $Q_1$ in (\ref{quant2}). With this aim, for a given finite
value of $a$ we take $\vert Q_2 \vert a<<1$. A straightforward calculation
shows that the second equation in (\ref{zeros}) uncouples into the system
\be 
Q_1 = \sqrt{v_0} \sin
(Q_1 a) \cosh (Q_2 a), \qquad Q_2 = \sqrt{v_0} \cos (Q_1 a) \sinh
(Q_2 a). 
\label{hiper} 
\ee 
In turn, the second one of these last equations reduces to 
\be 
\cos (Q_1a) \approx \frac{1}{a\sqrt{v_0}} =
\frac{1}{\eta}. 
\label{uncoup} 
\ee 
Now, let $\vert \delta \vert <<1$ be a correction of $Q_1$ around the
quantized values (\ref{quant2}), that is $Q_1 a \approx Q_{1,n}a +
\delta$. Then, equation (\ref{uncoup}) leads to
\be 
\delta_n = - \frac{\sin(\pi n/2)}{\eta},
\quad n \,\, {\rm odd}. 
\label{delta} 
\ee 
Thus, only odd values of $n$ are allowed into the approximation we are
dealing with. Now, to ensure small values of $Q_2$ we propose $Q_2a =
\sum_{m=1}^{+\infty} \lambda_m/\eta^m$. From (\ref{epsilon}) we have
\be
\frac{\epsilon_1}{v_0} \approx -1 + \left(\frac{aQ_{1,n}}{\eta}
\right)^2 - \frac{2 a Q_{1,n} \sin(\pi n/2)}{\eta^3} +
\frac{1-\lambda_1^2}{\eta^4} - \frac{2\lambda_1 \lambda_2}{\eta^5}
-\cdots 
\label{jota2} 
\ee 
In order to cancel the term including $\eta^{-4}$ we take $\lambda_1 = \pm
1$ (the appropriate sign will be fixed below). Higher exponents in the
power of $1/\eta$ are dropped by taking $\lambda_{m>1}=0$. In order to get
$\epsilon_1 \rightarrow E_n$, after introducing (\ref{quant2}) into
(\ref{jota2}), we realize that the condition $\eta >>1$ allows us to drop
the term including $\eta^{-3}$. Thus, at the first order in the
approximation of $Q_2a$ we get $\epsilon_1 \approx E_n$.

It is also necessary to have regard to the positiveness of energy
$\epsilon_1$, which is ensured whenever $n$ exceeds a minimum value. Let
us take $n:= n_{\inf} + m$, $m \in \mathbb Z^+$, where $n_{\inf}$ is the
ceiling function of $2\eta/\pi$, i.e. $n_{\rm inf} = \left\lceil
\frac{2\eta}{\pi}\right\rceil = \left\lceil \frac{2 a \sqrt{v_0}
}{\pi}\right\rceil $. Since $n$ is odd, the integer $m$ is even (odd) if
$n_{\rm inf}$ is odd (even). We finally arrive at
\be 
{\rm Re}(\epsilon) \approx \left( \left[ \frac{(n_{\rm inf}+
m)\pi}{2a\sqrt{v_0}} \right]^2 - 1 \right)v_0,  \quad m = \left\{
\begin{array}{ll}
0,2,4,\ldots & n_{\rm inf} {\rm \hskip1ex odd}\\[1ex]
1,3,5,\ldots & n_{\rm inf} {\rm \hskip1ex even}
\end{array}
\right. 
\label{quanti2} 
\ee 
On the other hand, it is worth noticing that the introduction of $Q_2a=\pm
1/\eta$ and (\ref{delta}) into the second equation in (\ref{epsilon})
leads to the conclusion that $Q_2a =+1/\eta$ has to be dropped. The
expression for the imaginary part of the complex energy $\epsilon$ is
calculated as follows. First, let us rewrite equations (\ref{hiper}) in
the form
\be 
K_1 = \pm \sqrt{v_0} \sin (Q_1a) \sinh(Q_2a), \qquad K_2 = \pm \sqrt{v_0}
\cos (Q_1a) \cosh(Q_2a). 
\label{hiper2} 
\ee 
Since $\epsilon_2 = -\Gamma/2$, from equations (\ref{epsilona}) we realize
that the sign of $K_1$ must be the opposite one of $K_2$. The introduction
of (\ref{uncoup}) into the second equation in (\ref{hiper2}) leads to
\be 
K_2 \approx \pm \frac{1}{a} \left( 1 + \frac{1}{2\eta^2} \right)
\label{k2} 
\ee 
where we have used $Q_2a \approx -1/\eta$. If $\eta >>1$ we get $aK_2
\approx \pm1$. Then, this last result together with the real part of the
factorization constant $\epsilon$ gives
\be 
K_1^2 = \epsilon_1 + K_2^2 \approx \epsilon_1 + \frac{1}{a^2}.
\label{k1} 
\ee 

\begin{figure}[ht]
\begin{center}
\includegraphics[height=.25\textheight]{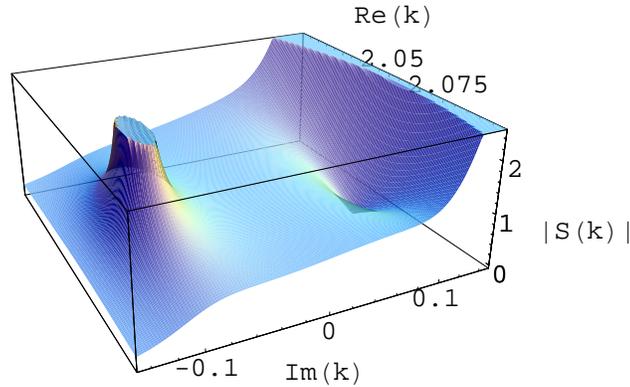}
\caption{\label{poloq1} \footnotesize The absolute value of the scattering
amplitude $S(k)$ close to the pole $k_{\alpha}$ (and the zero $\bar
k_{\alpha}$) which corresponds to the first resonance $E_{\alpha} =
k_{\alpha}^2$ of the square well reported in Table~\ref{tab1} for
$n_{\inf}=64$. The limits of our approach give the number $k_{\alpha}=
2.063412 -i\,0.099882$.}
\end{center}
\end{figure}

\noindent
Thus, for $a^2 \epsilon_1>1$ we obtain $K_1 \approx \pm
\sqrt{\epsilon_1}$. Finally, the introduction of (\ref{k2}) and (\ref{k1})
into the second equation of (\ref{epsilona}) produces
\be
{\rm Im}(\epsilon) = -\frac{\Gamma}{2} \equiv -2\vert K_1 \vert \,
\vert K_2 \vert \approx -\frac{2}{a} \sqrt{{\rm Re}(\epsilon)}.
\label{quanti4} 
\ee 
In Figure~\ref{poloq1} we show the absolute value of the scattering
amplitude close to one of its poles $k_{\alpha}$ and the corresponding
zero $\bar k_{\alpha}$. Some of the first resonances are reported in
Table~\ref{tab1} for different values of the potential strength $v_0$ and
the cutoff $a$.

\subsection{Complex Darboux-deformations of the radial square well}
\vskip6pt

\begin{figure}[ht]
\begin{center}
\includegraphics[height=.15\textheight]{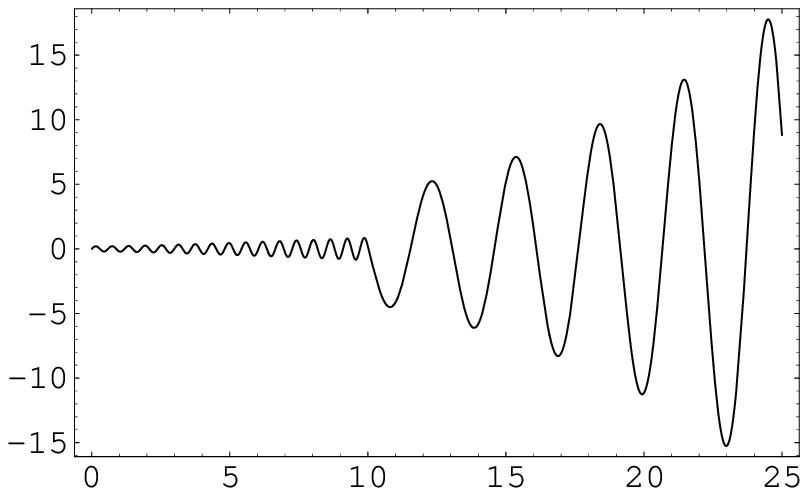} \hskip1cm
\includegraphics[height=.15\textheight]{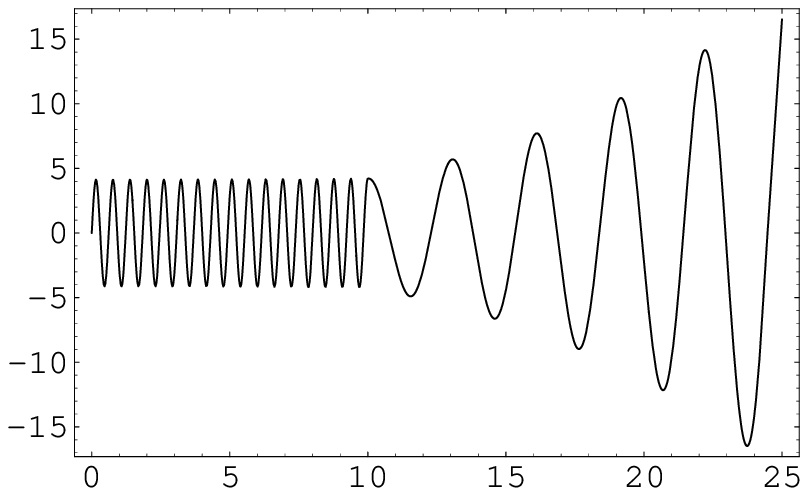}
\caption{\label{gamow} \footnotesize The real (left) and imaginary (right)
parts of the Gamow-Siegert function (\ref{gam1}) associated with the first
resonance state reported in Table~\ref{tab1} for $n_{\rm inf}=64$.}
\end{center}
\end{figure}

\begin{figure}[ht]
\begin{center}
\includegraphics[height=.15\textheight]{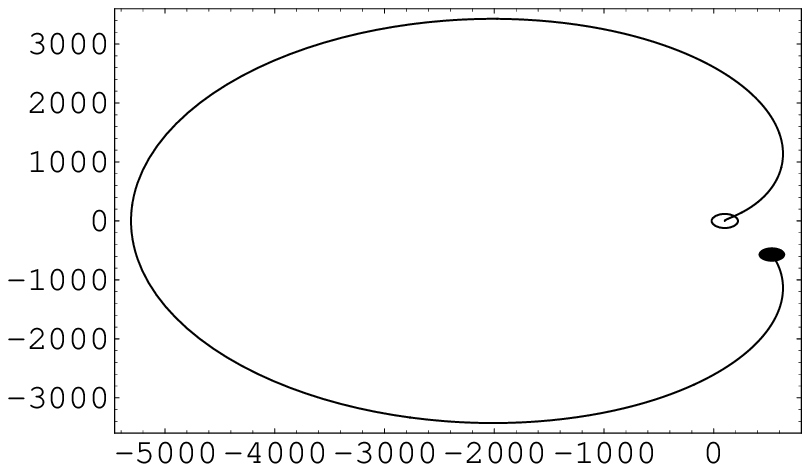} \hskip1cm
\includegraphics[height=.16\textheight]{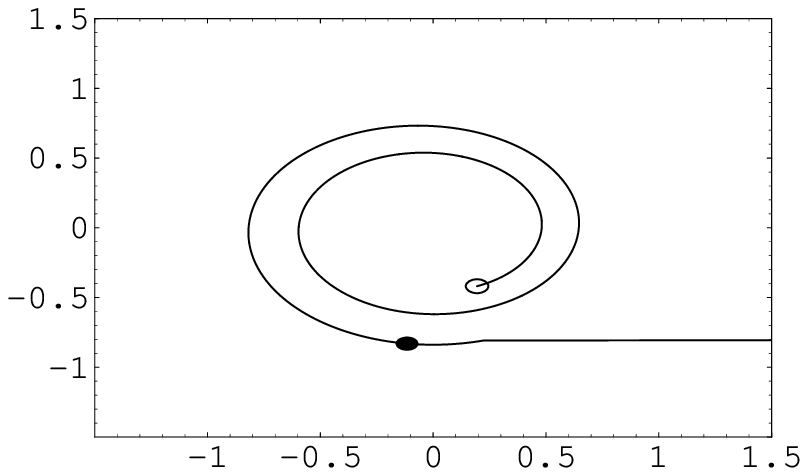}

\caption{\label{argand} \footnotesize The Argand-Wessel diagram of the
complex Darboux-deformed square well with $v_0=100$ and $a=10$. Left:
Detail of the cardiod-like behaviour of the new potential between $r=9.8$
(disk) and $r=10$ (circle). Right: The disk is evaluated at $r=10.1$ and
the circle at $r=13$. As complementary information: $\widetilde V_0(r=10)
= 100.2255 -i\,0.8076$ and $\widetilde V_0(r=10.1) =-0.1157 -i\, 0.8306$.}

\end{center}
\end{figure}

Figure~\ref{gamow} shows the global behaviour of a typical Gamow-Siegert
function associated with the radial square well. Notice the exponential
growing of the amplitude for $r>a$. This kind of solutions are used in
(\ref{Vtilde}) to complex Darboux-deform the square well as it is shown in
Figure~\ref{argand}.

The cardiod-like behaviour seems to be a profile of the complex potentials
derived by means of Darboux-deformations (compare with \cite{Ros07} and
\cite{Fer07b}). In this case, the complex potential shows concentric
cardiod curves for values of $r$ inside the interaction zone, one of them
is shown at the left of Figure~\ref{argand}. The real and imaginary parts
of $\widetilde V_{\ell}(r)$ are then characterized by oscillations and
changes of sign depending on the position in $r<a$. As a consequence,
these complex potentials behave as an optical device which both refracts
and absorbs light waves (see details in \cite{Fer07b} and discussions on
the optical bench in \cite{Mie04}). The presence of this kind of
oscillations is also noted at distances slightly greater than the cutoff
$r=a$. Hence, the complex Darboux-deformations (\ref{Vtilde}) are short
range potentials which enlarge the initial `interaction zone' as it is
shown in the right part of Figure~\ref{argand}.

\begin{table}[ht]

\begin{center}
\begin{tabular}{||l|l|l||}
\hline
\hline
    &$n_{\rm inf}=64$&$n_{\rm inf}=202$\\
\hline
$m=1$ & $04.247696 -i\, 0.412198$ & $16.791319 -i\, 0.819544$ \\
$m=3$ & $10.761635 -i\, 0.656098$ & $36.925312 -i\, 1.215324$ \\
$m=5$ & $17.472966 -i\, 0.836013$ & $57.256697 -i\, 1.513363$ \\
$m=7$ & $24.381689 -i\, 0.987556$ & $77.785474 -i\, 1.763921$\\
\hline
\end{tabular}
\end{center}

\caption{\label{tab1}{\footnotesize The first four resonance energies for
the radial square well of intensity $v_0 = 100$ and cutoff $a=10$ ($n_{\rm
inf}=64$) and for the radial square well of intensity $v_0 = 1000$ and
cutoff $a=10$ ($n_{\rm inf}=202$). Notice that, in each case, the even
values of $m$ obtained for the related one-dimensional problem are missing
(see e.g. Table~A.1 of reference \cite{Fer07b})}.}

\end{table}


\section{Concluding remarks}
\vskip6pt

We have studied the elastic scattering process of a particle by short
range, nonsingular radial field interactions. The poles of the involved
scattering amplitude $S(k)$ play a relevant role in the construction of
physical solutions connected with either bounded or scattering states of
the energy. It has been shown that poles $k_{\alpha}$ in the lower half
complex $k$--plane lead to the ``unphysical'' Gamow-Siegert functions
which are necessary to depict the behaviour of the scattering phenomenon.

The Gamow-Siegert functions were used to transform the radial short range
potentials into complex ones which preserve the initial energy spectrum.
These new potentials are `opaque' in the sense that they simultaneously
emit and absorb flux, just as an optical device which both refracts and
absorbs light waves. The transformation preserves the square-integrability
of the solutions at the cost of producing non-orthogonal sets of
wave-functions. It is also notable that scattering states are transformed
into scattering states while deformed Gamow-Siegert functions can be
either a new Gamow-Siegert function or an exponentially decreasing
function depending on the involved kinetic parameter $k$ and the pole
$k_{\alpha}$.

The method has been applied to the radial square well in the context of
$s$-waves ($\ell =0$). Analytical expressions were derived for the
corresponding complex energies in the long lifetime limit (i.e., for small
values of $\vert {\rm Im}(E_{\alpha})\vert =\Gamma/2$) and, as a
consequence, such energies fulfill a `quantization rule'. In contrast with
the `even' and `odd' resonances of short range potentials defined in the
whole of the straight-line (see e.g. \cite{Fer07b}), the resonances of
radial short range potentials are labelled by only odd (positive)
integers. This result enforces the interpretation of Gamow-Siegert
functions as representing quasi-bounded states: The bounded spectrum of a
potential which is invariant under the action of the parity operator
$v(x)=v(-x)$ includes odd and even functions. The symmetry is broken by
adding an impenetrable wall at the negative part of the straight line and
only the odd solutions are preserved. As we have shown, the same is true
for resonance states.

Finally, the results reported in this paper are complementary to our
previous work \cite{Fer07b}. It is remarkable that explicit derivations of
Gamow-Siegert functions are barely reported in the literature, not even
for simple models like those studied here (however see
\cite{del02,Ros07,Fer07a,Web82,Spr96,Ant01,Mad04}). We hope our approach
has shed some light onto the solving of the Schr\"odinger equation for
complex energies and functions fulfilling the purely outgoing boundary
condition.


\section*{Acknowledgments.}
The support of CONACyT project 24233-50766-F is acknowledged. The authors
are grateful to M. Gadella and D. Julio for interesting discussions.



\end{document}